\documentclass[a4paper,twoside]{article}

\usepackage{epsfig}
\usepackage{subcaption}
\usepackage{calc}
\usepackage{amssymb}
\usepackage{amstext}
\usepackage{amsmath}
\usepackage{amsthm}
\usepackage{multicol}
\usepackage{pslatex}
\usepackage{apalike}
\usepackage{algorithm2e}
\usepackage[bottom]{footmisc}
\usepackage{url}
\usepackage{booktabs}
\usepackage{pifont}
\usepackage{xcolor}
\usepackage{hyperref}
\usepackage{SCITEPRESS}     
\usepackage{setspace}
\usepackage{epstopdf}
\usepackage{todonotes}
\setuptodonotes{inline, bordercolor=white, backgroundcolor=yellow!21}

\newcommand\textvtt[1]{{\normalfont\fontfamily{cmvtt}\selectfont #1}}

\newcommand{\cmark}{\ding{51}}%
\newcommand{\xmark}{\ding{55}}%

\begin{document}

\title{Attestation with Constrained Relying Party}

\author{\authorname{Mariam Moustafa\sup{ 2, 3}, Arto Niemi\sup{1}, Philip Ginzboorg \sup{1} and Jan-Erik Ekberg\sup{1}}
\affiliation{\sup{1}Huawei Technologies Oy, Helsinki, Finland}
\affiliation{\sup{2}Aalto University, Espoo, Finland}
\affiliation{\sup{3}Denmark Technical University, Copenhagen, Denmark}
}




\abstract{Allowing a compromised device to receive privacy-sensitive
  sensor readings, or to operate a safety-critical actuator, carries
  significant risk. Usually, such risks are mitigated by validating
  the device's security state with remote attestation, but current
  remote attestation protocols are not suitable when the beneficiary
  of attestation, the relying party, is a constrained device such as a
  small sensor or actuator. These devices typically lack the power and
  memory to operate public-key cryptography needed by such protocols,
  and may only be able to communicate with devices in their physical
  proximity, such as with the controller whose security state they
  wish to evaluate. In this paper, we present a remote platform
  attestation protocol suitable for relying parties that are limited
  to symmetric-key cryptography and a single communication channel. We
  show that our protocol, including the needed cryptography and
  message processing, can be implemented with a code size of 6 KB and
  validate its security via model checking with the ProVerif tool.}

\onecolumn \maketitle \normalsize \setcounter{footnote}{0} \vfill

\section{\uppercase{Introduction}}
\label{sec:introduction}

Remote attestation allows a device to validate the
security state of another device \cite{Cok11}. A trustworthy mechanism, usually
with hardware backing, measures the security state of the device
where it resides and generates cryptographically protected attestation
evidence. The evidence constitutes a security proof that can be
remotely appraised by a verification service, typically by
matching measurements reported in the evidence against trusted reference
values. The result is a verdict that a relying party can then use as the basis of a trust
decision. Securely transmitting these messages -- the
attestation challenge, evidence and results -- is the task of a remote
attestation protocol. The protocol must provide at least message
integrity, freshness and origin authentication. This is usually
accomplished with public-key cryptography, such as asymmetric
signatures. Critical is also the protocol's resistance against relay 
attacks, where an attacker uses evidence generated by a valid device
to attest a compromised device.

After around two decades of research, remote attestation is now deployed commercially on servers and high-end consumer devices such as smartphones and PCs \cite{Nie23}. Standardization is also progressing, holding promise of interoperability in the future \cite{rfc9334}. On constrained devices, remote attestation is still
rarely used, despite an abundance of proposals from academia
\cite{Joh21} and industry \cite{diceAttArch,Hri22}. This stems from the constrained devices' lack of computing power and memory, which makes them incapable of performing the public-key cryptography required by most remote attestation protocols.
Research on resource-constrained attestation focuses on attesting the constrained device: 
protocols based on PUFs \cite{Sad11},
one-time signatures \cite{Rom23} and even statistics \cite{Neu20} have been proposed for this use case.

\begin{figure}[!ht]
\centering
\includegraphics[scale=0.6]{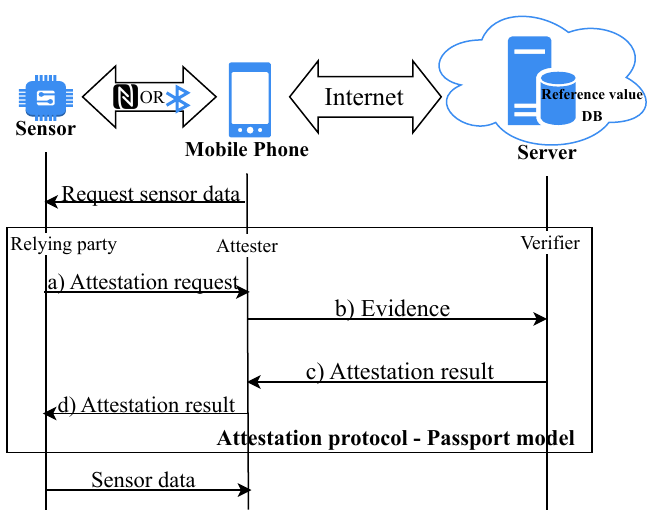}
\caption{Example of a attestation with a constrained relying party: a sensor validating the security of a mobile phone as a precondition for transmitting sensor readings. The example uses the passport model for communication in a remote attestation, where no direct communication between RP and verifier is needed.}
\label{fig:motivating-case}
\end{figure}

We consider a similar setup
but with a reverse attestation requirement: how can a constrained device confirm the
security state of the system it is interacting with? A sensor may measure data that reveals information of the user's
illnesses or lifestyle: releasing such information to another device without first validating the integrity
of the receiving software may result in a privacy violation. Similarly, a key tag should not transfer key
material to a malware-infested smartphone. In these kinds of settings, the constrained device is a relying party for attestation. A visualization of this is shown in Fig. \ref{fig:motivating-case}.

As the definition of ``constrained'', we work under the assumptions that the relying party device a) does not
support public-key cryptography; and b) can only communicate with an external attestation verifier via the attester, which makes man-in-the-middle attacks on protocol messages a specific concern. Such a network setup is dominant today --- in home
IoT Zigbee / BT-LE systems all radio communication is orchestrated via gateways, which 
in such systems are also the obvious targets of attestation. In personal area networks (PANs) the same applies, and a mobile phone takes the place of 
the gateway. Examples of PAN devices are phone headsets, heart monitors and blood-sugar sensors. 

The IETF RFC 7228 standard \cite{rfc7228} categorizes constrained devices into classes based on the amount of memory they have for data and code. The smallest ``sensor'' Class 0 devices have less than $10$ KB data and less than $100$ KB code; as asymmetric cryptography implementation typically requires several tens of kilobytes of code, there is no space for it, even if performance and battery consumption concerns are ignored. 

We present to our knowledge the first remote attestation protocol that neither uses public-key cryptography in the relying party, nor requires a second (trusted) channel for authentication/attestation. Our contributions are as follows:

\begin{enumerate}
    \item We analyze a less-considered attestation flow, where constrained devices need to attest a system before becoming part of its 
    operation.

    \item We provide an attestation protocol design where the relying party does not need public-key cryptography and can participate via a single insecure channel it has with the attester.

    \item We validate the security of our protocol using  formal model checking and provide a working proof-of-concept, whose performance metrics confirm the protocol's viability in practice. 
\end{enumerate}

\section{\uppercase{Background}}\label{sec:background}

\subsection{Remote Attestation}
\label{sec:passport model}

Following the IETF and Trusted Computing Group attestation architectures \cite{rfc9334,diceAttArch}, remote attestation protocols involve three active participants: \emph{attester}, \emph{verifier} and \emph{relying party} (RP). The RP wants to know the state of the attester before making a trust decision, such as granting access to a resource. The attester is equipped with a trustworthy mechanism \cite{Cok11} such as a Trusted Execution Environment (TEE) \cite{Gun22}\footnote{A TEE is an isolated computing environment in a device. It contains trusted  applications that process and store data independently from the device’s main operating system.}, or a Trusted Platform Module (TPM) \cite{Seg17}, which collects and cryptographically protects attestation evidence. The evidence may, for example, include boot-time code measurements or whether the current user has been authenticated.

The {\em verifier} appraises the evidence and issues a verdict (attestation results). 
The RP and verifier roles may be combined and implemented on the same device -- this has been common practice in academic work on attestation protocols \cite{Nie22b}. In the industry, complexity of evidence appraisal -- which involves acquiring trusted reference values and matching them against the measurements reported in the evidence -- has triggered a shift towards implementing the verifier as a separate online service. This is the more useful setup in the context of constrained devices: separating the two roles reduces the required complexity and code size in the RP device.

Two interaction patterns are commonly used in remote attestation: the background check and the passport model \cite{rfc9334}. The latter, illustrated in Fig. \ref{fig:motivating-case}, is better suited to resource-constrained RP, as it requires only a single communication link in the RP. The attester carries the attestation results produced by the verifier as a ``passport'', which the attester stores and then presents to the RP when needed. The attester may also obtain fresh results from the verifier if the current results are older than what the RP accepts.

\subsection{Related Work}\label{subsec:related-work}
We will next describe how previous remote attestation protocols cater to constrained devices, focusing on protocols based on symmetric-key cryptography.

The SlimIoT protocol \cite{Amm18} performs swarm attestation where the entities being attested are a ‘swarm’ of constrained IoT devices. The verifier broadcasts two challenges in a sequential manner which are stored by the attesters. Then the verifier discloses the keys used to generate the challenges so that the attesters can verify the stored challenges. If the keys are verified, the attesters generate the attestation evidence, aggregates the evidence from other attesters, and forwards the evidence. 

SCAPI \cite{2017scapi} is another swarm-based attestation protocol. It assumes that the attesters contain a Trusted Execution Environment (TEE) to execute tamper-resistant tasks. The verifier issues an attestation request to a specific device which in turn generates an attestation report and requests reports from other attesters in the network. Upon receiving the reports, the attester merges them and sends them to the verifier. Each attester has shared keys with all other attesters in the network which adds memory footprint and power consumption overhead. 

J{\"a}ger et al.~propose a protocol \cite{Jag17} where the server sends a nonce as a challenge, the attester hashes the nonce with the key being attested, and the verifier checks that the hash value is as intended. The protocol essentially attests key possession and does not consider attestation metrics where additional claims about the device are included. 

AAoT \cite{2018AAoT} is an attestation protocol based on physical unclonable functions (PUFs) -- special hardware that is used to generate the symmetric keys between the attester and verifier. The first stage of the protocol includes mutual authentication where the verifier and attester generate keys based on their identities using PUFs and performing a MAC based on their previously exchanged messages. If the MAC digests do not match, the protocol is aborted. Otherwise, the attester checksums the entire memory content in the device including the PUF component.

In the SIMPLE protocol \cite{Amm20}, the verifier generates a nonce, the value of a valid software state, and the MAC of these values. The attester verifies the MAC, computes its own state, and checks whether the computed state matches that the verifier sent. The attester sends the results of its check to the verifier to validate.  

All in all, these protocols cater to the case where the attester, instead of the relying party (RP), is the constrained device. This is evidenced by the lack of separation between verifier and RP roles. When the RP and verifier roles are both implemented on the same constrained device, only simple attestation metrics and appraisal procedures are possible, since verification is constrained by the resource limits of the RP device. Another drawback of some of the protocols is that they require synchronized clocks, or special hardware, like PUFs, in the constrained device.

\section{\uppercase{Requirements}}
\label{sec:Requirements}
We list below the functional and security requirements for the remote
attestation protocol with a constrained relying party.

\subsection{Functional Requirements}

To ensure the viability of our protocol in the use case where the
relying party (RP) is a constrained device, such as a class $0$ or
class $1$ sensor or actuator, we require the following:

\begin{itemize}
\item[FR1.] The RP can be implemented with $<10$ KB of code, including
  cryptography, but excluding the transport protocol such as UDP or
  Bluetooth.
\item[FR2.] The RP does not need public-key cryptography.
\item[FR3.] The RP only needs to communicate with the attester.
\end{itemize}

\subsection{Security Requirements}
\label{sec:adversary-model}
\label{sec:security-requirements}
We assume the Dolev-Yao attacker model: the attacker can read, intercept,
insert, relay and modify protocol messages, but is not able
to guess secret keys or to break cryptographic primitives \cite{Dol83}. The
attacker can use uncompromised devices as oracles and pretend to be
any of the participating entities, but is not able to compromise the
verifier or the TEE of the attester.

The attacker's goal is to trick an uncompromised RP into trusting a
compromised device. The uncompromised RP is assumed to execute its
part of the protocol correctly, so it will only trust an
attester after receiving a valid attestation result that the RP believes to
describe the security state of that particular attester. Thus, we 
formulate our security requirements in terms of the security of the
attestation results:

\begin{itemize}
\item[SR1.] Freshness of attestation results: the RP can detect whether an attestation result was generated in a particular run of the protocol.
\item[SR2.] Binding of attestation results to a particular attester: the RP can detect whether the verifier generated the result based on its appraisal of a particular attester.
\item[SR3.] Integrity of attestation results: the RP can detect whether an attestation result has been generated by a particular verifier,
  and whether the result has been modified in transit.
\item[SR4.] Confidentiality of attestation metrics and attestation results: the attester measurements and results are encrypted to ensure privacy. 
\end{itemize}

\section{\uppercase{Protocol Design}}
\label{sec:protocol-design}
Our protocol uses the passport model
discussed in Section \ref{sec:background}, with the provision that
attestation results cannot be reused and are bound to a
particular relying party. Accordingly, we call our protocol {Attestation Protocol for Constrained Relying Parties -- Live
  Passport Model}, or APCR-LPM for short. 
  
\begin{figure*}[!ht]
\centering
\includegraphics[scale=0.7]{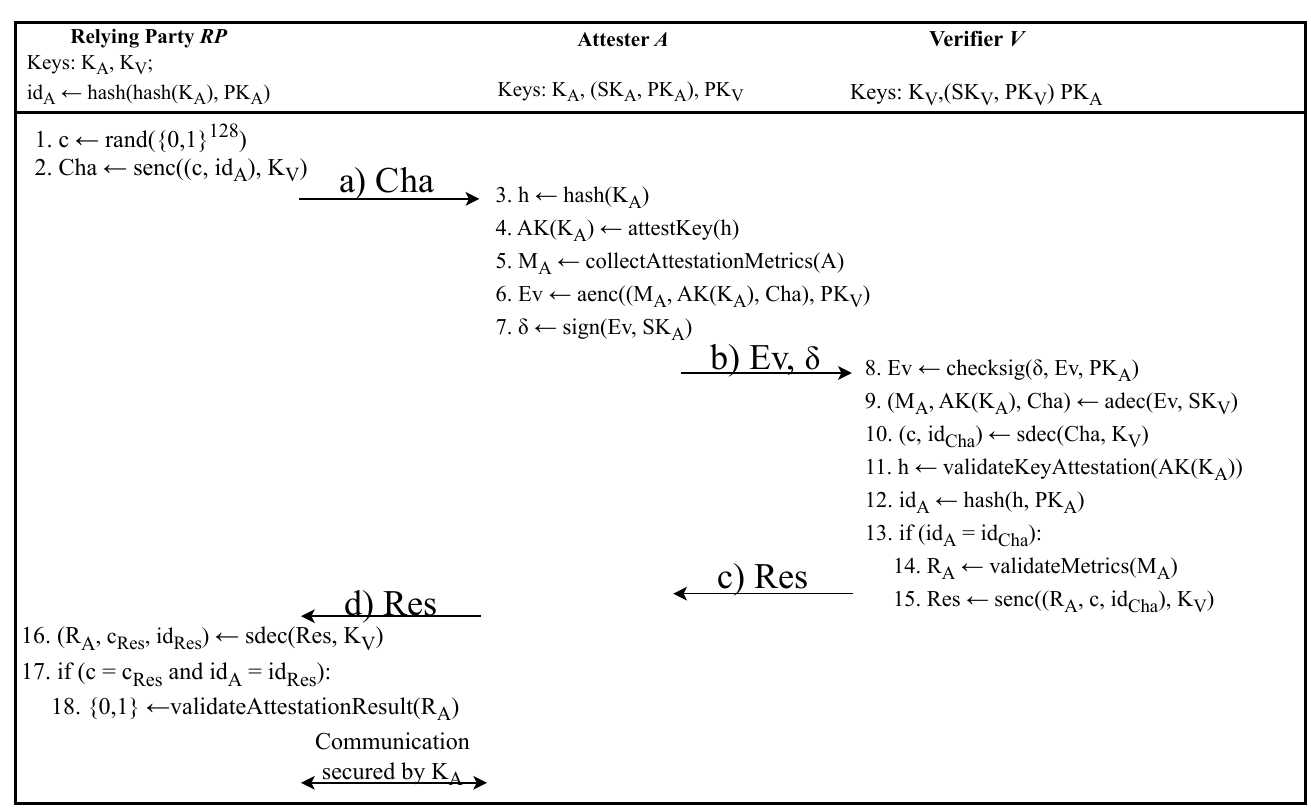}
\caption{Attestation Protocol for Constrained Relying Party.}
\label{fig:apcr-protocol}
\end{figure*}

Fig. \ref{fig:apcr-protocol} illustrates the
protocol steps and Table \ref{tab:notation} describes
our notation. For symmetric encryption (denoted $senc$ and $sdec$), we use a cipher that provides authenticated encryption, such AES-CCM. Thus the $sdec$ operation either returns the decrypted plaintext or, if integrity violation was detected, an error. Similarly, signature validation ($checksig(sig, m, PK)$) either returns the signed message or an error. The $aenc$, and $adec$ are encryption and decryption operations of a public-key authenticated encryption scheme, e.g., ECIES \cite{ts33501-2023}.

The protocol involves three principals:

\begin{itemize}
    \item Relying Party RP: a constrained device with little memory, supporting only symmetric-key cryptography. It has a communication link with the attester and can verify attestation results, but not attestation evidence.
    \item Attester A: a non-constrained device, such as a smartphone, that wants to prove its trustworthiness to the relying party. It has a trustworthy mechanism for evidence generation and secure storage for secrets, (often denoted a ``Trusted Execution Environment'' (TEE)).
    \item Verifier V: a non-constrained device, such as a cloud server, that is trusted by both RP and attester. The verifier can validate and appraise the evidence sent by the attester. RP and verifier are assumed to have agreed upon an evidence appraisal policy.
\end{itemize}

\textbf{Bootstrapping}. At the start of the protocol, the relying party (RP)
has symmetric keys $K_A$, and $K_V$, shared with the attester and verifier, respectively. 
We envision three possible key distribution scenarios:
\begin{enumerate}
\item The relying party (e.g., wearable, smart device) and attester belong to the same vendor. In these cases, the key material can be installed in these devices during manufacturing.
\item A user-based bootstrapping or pairing protocol involving an out-of-band channel has been executed by the owner of the devices,  which results in a shared key between devices.
\item The relying party (sensor) has a predefined relationship with the verifier. The verifier also takes on the role of a key distribution center that  creates and distributes session keys to the relying party and attester. We give a variant of our protocol for this scenario in the Appendix. 
\end{enumerate} 

The attester has an asymmetric keypair $(SK_A,PK_A)$ that is uses to authenticate itself to the verifier. The verifier is assumed to either trust $PK_A$ directly or able to construct a trust chain, e.g. with X.509 certificates, that allows it to trust $PK_A$. 
The RP also has an
identifier $id_A$ for the attester $A$ that is a function of both $K_A$ and $PK_A$.

 
\begin{table*}
\caption{Summary of notation.}
    \centering
\begin{tabular*}{\textwidth}{lll}
\hline
\textbf{Term} && \textbf{Description}\\
\hline
$K_A$ && Shared symmetric key between $A$ and $RP$.\\
$K_V$ && Shared symmetric key between $V$ and $RP$.\\
$(SK_V, PK_V)$ && The (secret key, public key) pair of $V$.\\
$(SK_A, PK_A)$ && The (secret key, public key) pair of $A$.\\
$h$ && Hash of $K_A$.\\
$c$ && A 128-bit pseudorandom value.\\
$M_A$ && The attestation metrics produced by $A$. \\
$r \leftarrow rand(\{0,1\}^{128})$ && Generate pseudorandom 128 bits string.\\
$c \leftarrow senc(m, K)$  && authenticated encryption of $m$ with shared key $K$.\\
$m \leftarrow sdec(c, K)$ && authenticated decryption of $c$ with shared key $K$.\\
$c \leftarrow aenc(m, PK)$  && public-key authenticated encryption of $m$ with public key $PK$.\\
$m \leftarrow adec(c, SK)$ && public-key authenticated decryption of $c$ with secret key $SK$.\\
$sig \leftarrow sign(m, SK)$ && Signing of $m$ with secret key $SK$.\\
$m \leftarrow checksig(sig, m, PK)$ && Verifying the signature of $m$ with key $PK$. \\
$h \leftarrow hash(m)$ && Computing the hash of $m$.\\
$AK(K) \leftarrow attestKey(h)$ && Attestation of key $K$ by TEE.\\
$h \leftarrow validateKeyAttestation(AK(K))$ && Validate that key $K$ is attested by TEE.\\
$M \leftarrow collectAttestationMetrics(E)$ && Compute the attestation metrics of entity $E$.\\
$R \leftarrow validateMetrics(M)$ && Compute attestation results $R$ based on the metrics $M$.\\
$\{0,1\} \leftarrow validateAttestationResult(R)$ && Determine trustworthiness based on the attestation results $R$.
\end{tabular*}
    \label{tab:notation}
\end{table*}
The protocol steps, shown in Fig. \ref{fig:apcr-protocol} are as follows: 
(1) The relying party (RP) prepares a challenge by generating the nonce $c$. It encrypts $c$ and the identifier $id_A$ of the attester using $K_V$ (2). The nonce $c$ acts as a session identifier as well as a freshness value. RP sends the challenge to the attester A, message (a). The attester cannot read the contents of the message as it is encrypted with a key that it does not know.

 The attester's TEE performs key attestation on the hash of $K_A$ (4). By including the key attestation, the TEE vouches that $K_A$ cannot be extracted from the TEE. In step (5), the attester collects the attestation metrics of the device and its software. Then, it encrypts the key attestation, collected metrics, and challenge it received from RP using $PK_V$ (6), signs the encrypted evidence using its private key $SK_A$ (7), and finally sends the evidence and signature to the verifier in message (b).

 The verifier verifies the signature of the evidence using $PK_A$ (8) and decrypts the evidence with $SK_V$. Then it decrypts the challenge to extract $c$ and $h$ (10). In step (11), it verifies the key attestation and computes its own version of the attester’s $id_A$ based on the public key it has used for checking the signature (12). In step (13), the verifier checks whether $id_A$ received in the challenge is equal to that it has computed. If any of those steps fail, the verifier aborts the protocol. Based on the metrics, the verifier generates a verdict or attestation result for RP (14) and encrypts the attestation results together with the nonce $c$ and $id_{Cha}$ using $K_V$ (15) resulting in the ciphertext $Res$. The verifier sends $Res$ to the attester in message (c). The attester 
forwards $Res$ to RP in message (d). The value $id_{Cha}$ is included in Res as an indicator to RP that the verifier has verified the evidence of the attester with identity $id_{Cha}$. It also includes the decrypted challenge ($c$, $id_{Cha}$) to maintain the freshness of the message and to indicate to RP that the intended verifier has received the challenge and decrypted it. 


(16) RP decrypts $Res$ and checks that the values of $c_{Res}$ and $id_{Res}$ are equal to the ones it sent in the challenge (17). RP can then process the attestation result to determine the attester's state (18).

Subsequent, application-specific communication between RP and attester  depends on the result of step (18); this communication is secured using $K_A$.

\section{\uppercase{Security Analysis}}

We analyze the security of ACPR-LPM first via informal discussion  and then formal model checking.

\subsection{Discussion}

APCR-LPM fulfills the security requirements
of Section \ref{sec:security-requirements} as follows:

\textbf{SR1} (Freshness of attestation results). The relying party
(RP) includes a nonce $c$ in the challenge and the verifier is
required to include the same nonce in the attestation results. The
nonce is a pseudorandom number that is generated fresh in every run
of the protocol. In step (17), the RP checks whether the nonce in
the received attestation result matches the nonce it sent in the last
challenge. For a replayed result, the check will fail. The RP could
also use a protocol timeout to prevent the acceptance of obsolete results.

\textbf{SR2} (Binding of attestation results to a particular
attester). The RP binds the challenge $Cha$ to the identity of a
particular attester by including the value $id_A = hash(hash(K_A),
PK_A)$. The verifier knows $PK_A$, the public key of the attester's
TEE, and receives $hash(K_A)$ from the key attestation included in the
attestation evidence, so it can compute a reference $id_A$. By
verifying the evidence signature with $PK_A$ in step (8) and by
comparing the self-computed $id_A$ against the one decrypted from
$Cha$ in step (13), the verifier can detect whether the evidence was
generated by a TEE that the verifier trusts and that belongs to the
attester the RP intended. Since we assume the evidence signing keys
($SK_A$) to be unique to the TEE instance and unextractable, the verifier
can validate that the evidence was generated by the particular TEE
that is identified with $PK_A$. The verifier includes $c$ and the
validated $id_A$ in the results, allowing the RP to check, in step (17),
that they match the values it sent in the challenge. This fulfills the
requirement, preventing relay attacks, sometimes called Cuckoo attacks
\cite{2008cuckoo,2020relayexplanation}, a common issue with remote
attestation protocols \cite{Nie21,Ald23}.

\textbf{SR3} (Integrity of attestation results). The RP checks the
integrity of the attestation result by decrypting, in step (16), the
result message $Res$ with the shared key $(K_V)$ it has with the
verifier. Since $Res$ is protected with authenticated encryption, using
a key ($K_V$) that the attacker does not know, the RP can detect
whether the message was modified after encryption or generated by
a different verifier. This fulfills the
integrity requirement. Finally, in step (18), the RP can evaluate the
trustworthiness of the attester it identified in the challenge by
examining the verifier's verdict that it decrypts from $Res$.

\textbf{SR4} (Confidentiality of attestation metrics and attestation results). The confidentiality of the attestation metric is guaranteed with public key cryptography, where the evidence is encrypted using the verifier’s public key $PK_V$ in step (6). Only the verifier can read the evidence using its private key $SK_V$. The attestation result, on the other hand, is encrypted using the symmetric key $K_V$ the verifier shares with the relying party, step (15). Only the parties who know $K_V$ can read the attestation results. 

\subsection{Formal Model Checking}\label{subsec:model-checking}

We used the ProVerif tool \cite{2018proverif} to formally model our protocol and verify its security properties. 
%
Queries describing the desired security properties are included in the ProVerif model. These queries are written in terms of ProVerif events which mark certain stages reached by the protocol and have no effect on the actual behavior of the model. The tool attempts to explore all possible execution paths of the protocol, trying to find a path where a query fails. ProVerif assumes the Dolev-Yao attacker model and can perform replay, man-in-the-middle and spoofing (impersonation) attacks, which aligns with the adversary model in Section \ref{sec:adversary-model}. Our ProVerif code is available in GitHub \footnote{\href{https://github.com/Mariam-Dessouki/RAforConstrainedRP/tree/apcr-paper}{\color{blue}{ProVerif Model}}}. 

The following query represents the security requirements SR1, SR2, and SR3 in Section \ref{sec:security-requirements}:
{\scriptsize
\begin{spacing}{0.5}
\begin{equation*}
\begin{aligned}
& query\ PK_A: pkey,K_A: key, K_V: key, R_A: bitstring, c: nonce, \\
& h: bitstring, id: bitstring, M_A: bitstring, Cha: bitstring; \\
& inj-event(relyingPartyAccepts(K_V, R_A, c, id)) \\
& \implies  inj-event(relyingPartyBegins(K_V, c, id))\ \&\& \\
&\ \ \ \ \ \ \ \ inj-event(attesterBegins(PK_A, h, M_A, Cha))\ \&\& \\
&\ \ \ \ \ \ \ \ inj-event(verifierAccepts(PK_A, K_V, M_A, id, c))\ \&\& \\
&\ \ \ \ \ \ \ \ R_A = validateEvidence(M_A)\ \&\& \\
&\ \ \ \ \ \ \ \ id = hash((h, PK_A))\ \&\& \\
&\ \ \ \ \ \ \ \ Cha = senc((c,id), K_V).
\end{aligned}
\label{def:rp-query}
\end{equation*}
\end{spacing}
}

The query defines an injective correspondence between the event of the relying party (RP) accepting $R_A$ and all other events in the protocol. This means that for each occurrence of the event $relyingPartyAccepts$ there is a distinct occurrence of all other events in the query. The RP will only accept the protocol run if there has been a previous run where it:
\begin{enumerate}
    \item initiated the protocol by sending the encrypted $c$ and $id$ ($relyingPartyBegins$);
    \item the attester has accepted this encrypted $c$ and $id$ as $Cha$ and collected attestation metrics $M_A$ ($attesterBegins$);
    \item the verifier has received $M_A$ and decrypted $Cha$ ($verifierAccepts$).
\end{enumerate}
There is an added constraint on the relation between the metrics $M_A$ and the attestation results $R_A$. This query models the security requirements (Section \ref{sec:security-requirements}) by including $c$ in the events for freshness, the keys for data origin authentication and $M_A$ and $R_A$ for integrity. The $M_A$ is matched to both the attester and verifier event and the RP will not accept the final message unless $R_A$ is a function of $M_A$ and it has received back the encrypted $c$ and $id$ it used in the first message.

In order to satisfy SR1 (freshness of attestation results), the RP subprocess simulates the protocol by generating a new nonce $c$ with every protocol run.

When the RP receives the value $Res$, it does the following.
{\small
\begin{verbatim} 
let (R_a:bitstring, =c , =id)=sdec(Res, K_v) in 
event relyingPartyAccepts(K_v, R_a, c, id);
\end{verbatim}
}
The RP first checks that the nonce $c$ is the same as the one it has sent and then it invokes the event $relyingPartyAccepts$. The equal sign (before $c$ and $id$) matches the decrypted value to an already defined value. If the received nonce does not match it an error would occur and the event would not be invoked. For SR3 (integrity of attestation result), the RP checks the integrity of the received $Res$ by using the key $K_V$ it shares with the verifier to decrypt the message. If the values of the received $c$ and id are not equal to the sent values then the event will not be invoked.  

The following code checks the binding of the attestation result to a particular attester (SR2). It is executed after the verifier checks the signature of the evidence.
{\small
\begin{verbatim}
  let id = hash((h, PK_a)) in
  if (id = id_cha) then
   let R_a = validateEvidence(M) in
   event verifierAccepts(PK_a, K_v, M, id, c);
\end{verbatim}
}

The event $verifierAccepts$ is invoked after the verifier subprocess checks that the $id$ value received from the RP matches the $id$ value it generated from the attester’s public key, otherwise $verifierAccepts$ would not occur.

The query would fail if any of the security requirements are not satisfied. ProVerif was able to check all possible protocol states and terminate. It did not find an attack against the query defined above, i.e, the security requirements SR1, SR2 and SR3 are satisfied. 

The query used to represent the security requirement SR4 is as follows:
{\small
\begin{equation*}
\begin{aligned}
& query\  K_V: key, id: bitstring, R_A: bitstring, c:nonce; \\
&attacker(R_A)\ \&\&\ event(relyingPartyAccepts(K_V, R_A, c, id)) \\
& \implies false.
\end{aligned}
\label{def:secrecy-query}
\end{equation*}
}
The query states that the events of the attacker knowing the attestation result $R_A$ and the RP accepting the same attestation result cannot occur together. 
Note that although it is not explicitly mentioned in the query, the query also includes the secrecy of the attestation metrics $M_A$. 
The attestation results $R_A$ is a function of $M_A$, so even if $R_A$ is encrypted, an attacker that knows $M_A$ can easily derive  $R_A$ using the function $validateMetrics$ (see (14) in Fig. \ref{fig:apcr-protocol}). ProVerif did not find an attack on this query meaning that the attestation metrics and results are confidential, thus satisfying SR4.


\section{\uppercase{Proof-of-concept}}
To study the feasibility of our protocol on constrained devices, we implemented the relying party (RP) role
on the nRF5340 development kit \cite{2023nRF5340}. The RP was set up to communicate with a laptop over Bluetooth. Since the main advantage of the protocol is that it is suitable for constrained RPs, only the communication between RP and the attester (messages (a) and (d)) and the actions in the relying party (steps (1, 2, 16, 17, 18)) found in Fig.~\ref{fig:apcr-protocol} were implemented.

As an example use case we took 
the electronic lock-and-key system, where the user wants to use his mobile phone (instead of a NFC or Bluetooth keytag) to open doors --- in essence copying the key material normally residing on the keytag. The computationally weak keytag needs a way to attest the security of a mobile phone before releasing any cryptographic secrets to it. 

Our attestation protocol is applicable to this setup as follows. 
The keytag (relying party) initiates the attestation protocol over a near-field communication channel with the smartphone (attester), who in turn relies on a network server (verifier) to prove the security level of the TEE in the smartphone. After successful verification of attestation results, the keytag transfers the door key to the smartphone TEE.


\subsection{Board Setup} 
The nRF5340 board has a 64 MHz Arm Cortex-M33 CPU, 256 KB Flash and 64 KB RAM. It supports many interfaces including Bluetooth Low Energy and NFC. The manufacturer provides an SDK that includes the Zephyr RTOS (Real-Time Operating System), an operating system based designed for resource constrained devices. Zephyr can run on devices with 32KBs of RAM and has built-in support for multiple cryptographic libraries. For cryptography, we used primitives from the TinyCrypt library \cite{2017tinycrypt}, namely AES-CCM, PRNG and HMAC. 

\begin{figure}[!ht]
  \centering
\begin{subfigure}{0.45\textwidth}
         \caption{Message flow}
         \includegraphics[width=1\linewidth]{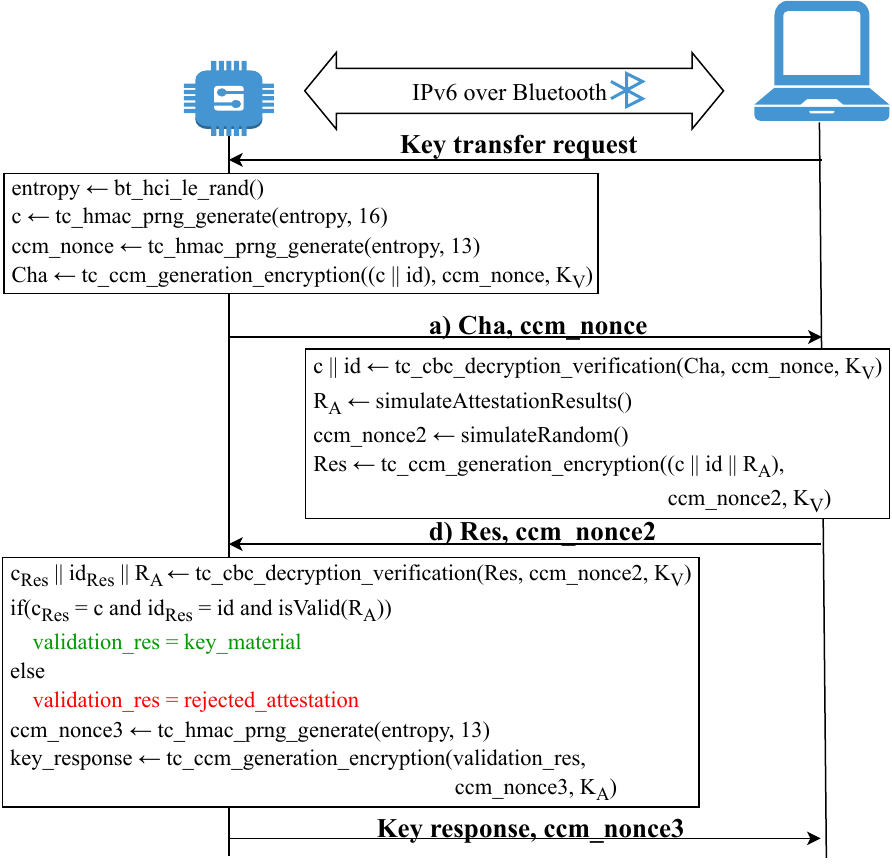}
\label{fig:prototype}
\end{subfigure}
\begin{subfigure}{0.4\textwidth}
         \caption{EAR object}
         \includegraphics[width=1\linewidth]{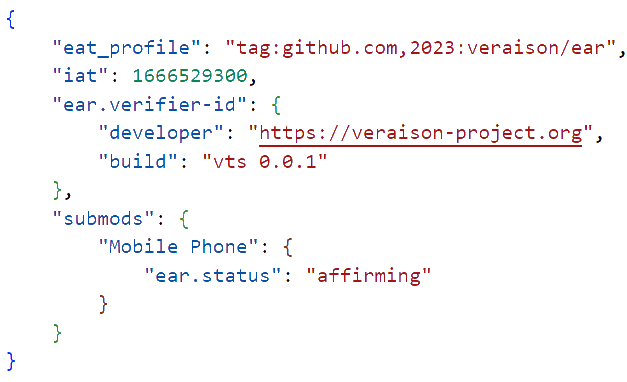}
         \label{fig:ear-object}
\end{subfigure}
\caption{Prototype implementation of keytag application.}
\label{fig:implementation}
\end{figure}

We extended Zephyr's sample application, called IPSP \cite{2023ipspsample},  with the implementation of the key transfer protocol shown in Fig. \ref{fig:prototype}. The connection between the relying party and attester in this prototype is IPv6 over BLE (Bluetooth Low Energy).

\subsection{Implementation}
As illustrated in Fig. \ref{fig:prototype}, after the Bluetooth connection is established, the laptop (attester) issues a key transfer request which initiates the ACPR-LPM protocol. The keytag (RP) generates a nonce and encrypts both the nonce $c$ and $id$ using $K_V$ and AES-CCM. The resulting ciphertext, $Cha$, is sent over the channel in message (a). Upon receiving that message, the application decrypts $Cha$ to extract the values $c$ and $id$. Then it encodes the attestation result object and sends the encrypted $c$, $id$, and $R_A$ over the Bluetooth channel. The attestation result object follows the Entity attestation token Attestation Result (EAR) \cite{2023eatarrfc} emerging standard. An example of an EAR JSON object is shown in Figure \ref{fig:ear-object}. The displayed required fields contain information about the EAR version, the issued at time, the verifier identity, the attester’s identity and the verifier’s verdict. The result object is encoded using the CBOR binary serialization format \cite{rfc8949} as CBOR has a small code footprint and encoding size. 
Depending on the $Res$ received in message (d), the keytag may decide to send the key material or not. In either case, the response has the same message size to mitigate side channel attacks. The current prototype does not implement all the steps in the attestation protocol; it is mainly concerned with implementing the relying party application.

\subsection{Measurements} \label{sec:measurements}
We compared the RP application against the original IPSP application to determine the increase in RAM and Flash requirements, shown in Table \ref{tab:rp-application-measurements}. The Flash memory increased by 6 KB and the RAM increased by less than 1 KB. This includes protocol processing, code from the TinyCrypt and zcbor libraries, and code specific to the keytag application. To exclude Zephyr from the measurements, we analyzed the object files and the size of the program was computed. The RP implementation, including protocol processing and code that calls TinyCrypt or zcbor, required around 2.3 KB of extra code. The total flash and memory size are the reported values Zephyr displays when building the application. The application code size was determined using the \textvtt{size} command-line tool on the RP application's object file.

The Zephyr ELF (Executable and Linkable Format) file was analyzed to determine the extra space the TinyCrypt and zcbor library have used. The command “nm –size-sort –radix=d zephyr.elf” was used to extract the sizes of the library functions used in the relying party application. Only the text size was measured as it is the application code being executed. Table \ref{tab:libraries-flash-size} shows the results. The application object file was analyzed in the same manner to determine the size of the different components of the code. The results are shown in Table~\ref{tab:application-component-size}.

\begin{table}[!ht]
        \centering
    \caption{Memory footprint of applications (in bytes).}
    \begin{tabular}{lll@{}}
    \toprule
    \multicolumn{1}{c}{}  
    \textbf{\begin{tabular}[l]{@{}l@{}}Memory \\ Type\end{tabular}} & \textbf{\begin{tabular}[c]{@{}l@{}}IPSP \\ Sample\end{tabular}} & \textbf{\begin{tabular}[c]{@{}l@{}}RP \\ Application\end{tabular}} \\ \midrule
    \textbf{Total Flash}                                                      & 241320                                                          & 247320 (+6000)                                                    \\
    \textbf{Total RAM}                                                        & 60280                                                           &  61184 \phantom{x}(+904)                                                     \\
    \textbf{\begin{tabular}[l]{@{}l@{}}Application \\ Code Size\end{tabular}} & 1792                                                            & 4396  \phantom{xx}(+2604)                                                      \\ \bottomrule
    \end{tabular}
    \label{tab:rp-application-measurements}
\end{table}

\begin{table}[!ht]
        \centering
        \caption{Libraries flash code size.}
        \begin{tabular}{@{}ll@{}}
        \toprule
        \textbf{Libraries Flash} & \textbf{Size [bytes]} \\ \midrule
        TinyCrypt                & 2304                  \\
        zcbor                    & 398                   \\ \bottomrule
        \end{tabular}
        \label{tab:libraries-flash-size}

\end{table}

\begin{table}[!ht]

    \centering
    \caption{Application components code size.}
    \begin{tabular}{@{}ll@{}}
    \toprule
    \textbf{Application code}                                            & \textbf{Size [bytes]} \\ \midrule
    \begin{tabular}[c]{@{}l@{}}Initialization \\ \& Logging\end{tabular} & 238                   \\
    Network Stack                                                        & 1606                  \\
    Cryptography                                                         & 272                   \\
    EAR decoder                                                          & 542                   \\ \bottomrule
    \end{tabular}
    \label{tab:application-component-size}

\end{table}

Overall, the “keytag” transfers 174 bytes: 1) 55 bytes $Cha$, which includes 16 bytes $c$, 16 bytes $id$, 13 bytes AES-CCM nonce, and 10 bytes MAC; and 2) 119 bytes encrypted key material in the last message. 
Reducing $Cha$ size to 16 bytes (one AES block) and analyzing the security implications is left for the future.

The ``keytag'' receives 194 bytes: 20 bytes key transfer request in the first message, and 174 bytes encrypted attestation results in message (d). 

To test the time overhead of APCR-LPM, we did three experiments, described below. In each experiment the timing of operations was repeatedly measured 10 times on the attester (laptop) side.

\begin{enumerate}
\item Baseline communication cost. We measured the time between attester sending a key request and it receiving a key response from the RP, without attestation messages (a) and (d), and without cryptographic operations in the RP. This takes 93 ms on the average. 
\item Protocol cost. We recorded the elapsed times (i) from when the attester sends a key transfer request to when it receives message (a), and (ii) from sending message (d) to it receiving key response and the ccm\_nonce3. The time required for the protocol steps in the attester was omitted. The sum of (i) and (ii), which is 289 ms on average, measures the cost of overall protocol communication and the cost of protocol processing in the resource-constrained RP device.
\item Protocol communication cost. We sent the same number and size of messages as in experiment (2), but omitted other processing such as message decoding and cryptography. The result was 281 ms on average, 7 ms less than the full protocol cost.
\end{enumerate}

We conclude that APCR-LPM has a time overhead of about 200 ms, compared to insecure communication with the RP. Most of the overhead is due to the additional round-trip, rather than encryption, decryption, and attestation result parsing in the RP.



\section{\uppercase{Results and Discussion}}
\begin{table*}[!ht]
\centering
\caption{Differences between APCR-LPM and other symmetric encryption-based RA protocols. The meaning of the columns is as follows: ``Separate RP'': the RP and verifier roles are separated. ``HW-independence'': the protocol can be implemented on any hardware. ``HW-based RA'': attestation uses a hardware-backed trustworthy mechanism. ``Universal attestation'': the protocol can be used to carry any attestation evidence format and arbitrary attestation claims instead of a small or fixed set.}
\small
\begin{tabular}{lcccc}
\hline
\textbf{Protocol} & \textbf{\begin{tabular}[c]{@{}c@{}}Separate\\ RP\end{tabular}} & \textbf{\begin{tabular}[c]{@{}c@{}}HW-inde-\\ pendence\end{tabular}} & \textbf{\begin{tabular}[c]{@{}c@{}}HW-based\\ RA\end{tabular}} & \textbf{\begin{tabular}[c]{@{}c@{}}Universal\\attestation\end{tabular}} \\ \hline
SlimIoT \cite{Amm18}          & \xmark                                                            & \xmark                                                               & \cmark                                                         & \cmark                                                                  \\
SCAPI \cite{2017scapi}            & \xmark                                                            & \xmark                                                               & \cmark                                                         & \cmark                                                                  \\
Rolling DICE \cite{Jag17}    & *                                                                 & \cmark                                                               & \cmark                                                             & \xmark                                                                  \\
AAoT \cite{2018AAoT}             & \xmark                                                                 & \xmark                                                               & \cmark                                                         & \xmark                                                                  \\
SIMPLE \cite{Amm20}            & \xmark                                                            & \cmark                                                               & \xmark                                                         & \xmark                                                                  \\
APCR-LPM [this paper]          & \cmark                                                            & \cmark                                                               & \cmark                                                         & \cmark                                                                 
\end{tabular}
\label{tab:related-work-summary}

\caption*{ (*) No RP but theoretically the roles of attester and verifier can be reversed so that verifier is a constrained device.}
\end{table*}
The APCR-LPM does not need asymmetric cryptography in the relying party, making it suitable for devices with limited resources. The measurements summarized in Section \ref{sec:measurements} show that the protocol is sufficiently small to be embedded in simple devices. 
According to the formal verification by ProVerif, the protocol is secure under an active Dolev-Yao attacker.

Previous remote attestation protocols for constrained devices, discussed in Section \ref{subsec:related-work}, either combine the relying party and verifier roles in the same device, with significant drawbacks in memory footprint and complexity, or require certain hardware in the constrained devices for the protocol to work. In practice, a relying party might be a network gateway, actuator, sensor, or keytag that may not be able to perform all the functionality that is requested of a verifier. Also, a protocol that does not require special hardware is more scalable. Table \ref{tab:related-work-summary} analyzes the differences between the earlier protocols and APCR-LPM. The separation of roles between the relying party and verifier in our protocol allows processing complex attestation evidence without increase in the memory footprint or processing power in the relying party. The relying party only needs to process attestation result, a standardized verifier's verdict on the evidence, following, e.g., the EAR proposal \cite{2023eatarrfc}.

A drawback of APCR-LPM is that it requires distribution of symmetric keys prior to protocol run. How to best achieve this depends on the relation between the parties. In the Appendix we describe a variant of the protocol that does not involve prior key distribution between the relying party and attester: the shared key is generated by the verifier and distributed to RP and attester during the protocol run. The flip side of this is that the verifier knows the key shared between the RP and the attester.

\section{\uppercase{Conclusion}}
\label{sec:conclusion}

We have designed and implemented APCR-LPM -- an attestation protocol for situations where the beneficiary of attestation, the relying party, is a very simple device, capable only of symmetric key operations and able to communicate only with the device whose trustworthiness it wants to evaluate. Our example use case is an electronic lock-and-key system, where the device generating the key is a dongle. In our implementation, the constrained device is an nRF5340 development board which connects via Bluetooth to a Linux laptop and performs relying party role in our protocol. The application requires 6 KB of code and 287 ms to perform the key transfer and attestation functionality. We used ProVerif to verify the protocol's security properties. In conclusion, the protocol is practical for the intended use cases.




\bibliographystyle{apalike}
{\small
\bibliography{allsources}}

\section*{\uppercase{Appendix}}
\begin{figure*}[t]
\centering
\includegraphics[scale=0.7]{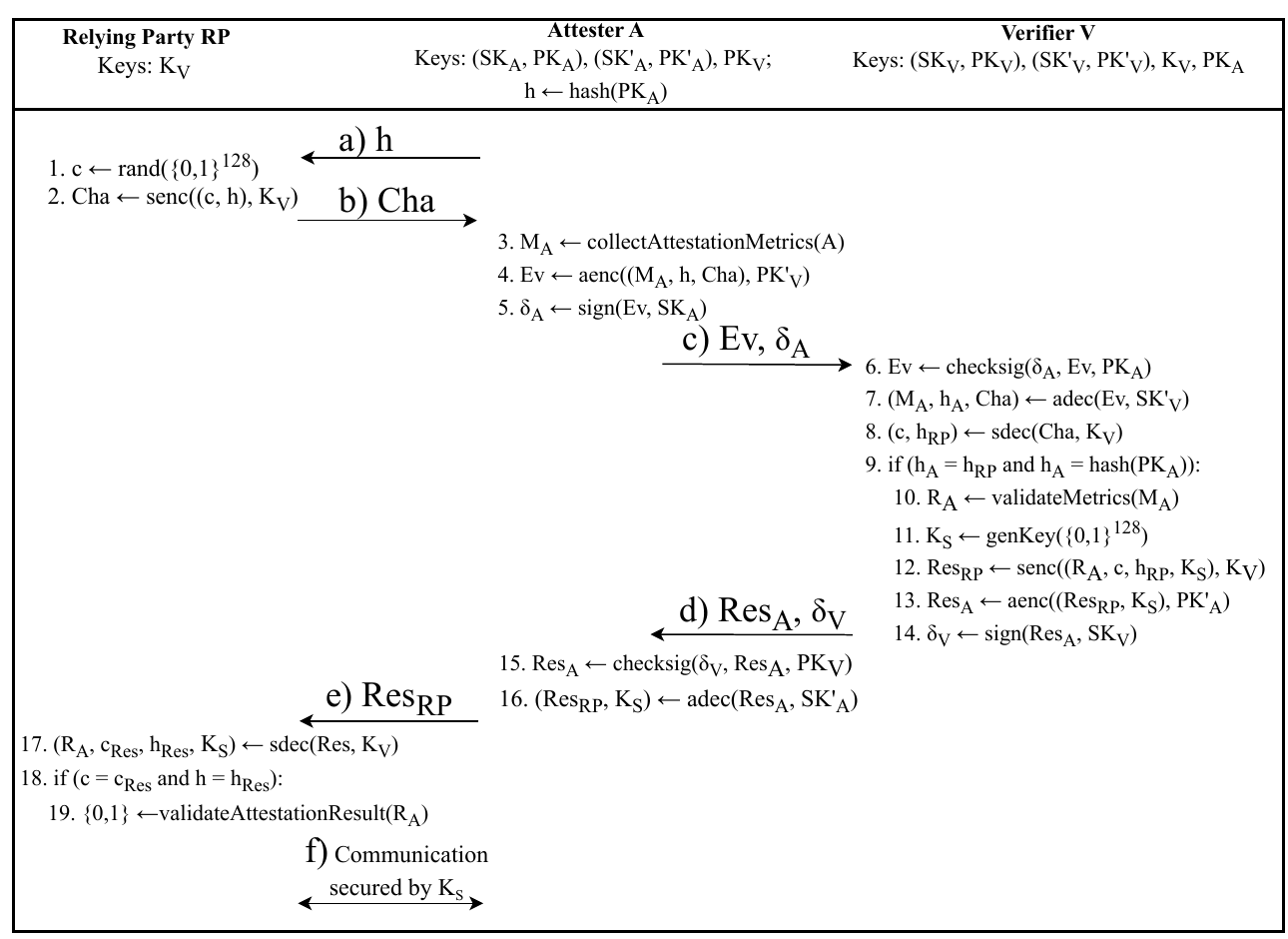}
\caption{Variation of APCR-LPM with no pre-shared keys between relying party and attester. 
}
\label{fig:verifier-generates-key}
\end{figure*}
We will now describe a variant of the protocol where the distribution of shared key to RP and attester is done  during the protocol run.

Initially, the relying party (RP) has a symmetric key $K_V$ that is unique to the RP device and shared with the verifier (V). 
The attester has two asymmetric keypairs $(SK_A,PK_A)$ and $(SK_A',PK_A')$ for signing and encryption, respectively. The hash of $PK_A$ is denoted by $h$. The verifier has two asymmetric keypairs $(SK_V,PK_V)$ and $(SK_V',PK_V')$, for signing and encryption, respectively. The verifier and the attester know each other’s public keys.

After the protocol run the RP has validated the result of the verifier's assessment of the attester's state; and the RP and the attester have a verifier-generated session key $K_S$. 

The protocol steps are as follows (cf. Figure \ref{fig:verifier-generates-key}). 
The attester sends to the RP message (a) containing the hash $h$ of its public key $PK_A$. (1) The RP prepares the challenge it will send to the attester by generating the nonce $c$. (2) It creates the challenge Cha by encrypting using $K_V$ the nonce $c$ and $h$, and sends Cha to the attester in message (b). The attester cannot decrypt Cha, because it does not have $K_V$.

(3) The attester collects the attestation metrics $M_A$ of the device, and (4) creates the evidence Ev, comprising of  $M_A$, $h$, and Cha encrypted with $PK_V'$. (5) It signs the Ev with its private key $SK_A$, obtaining the signature $\delta_A$, and sends the Ev and $\delta_A$ to the verifier in message (c).

(6) Upon receiving message (c), the verifier checks the signature $\delta_A$. If this check succeeds, it decrypts the Ev using the key $SK_V'$, and obtains $M_A$, $h_A$, and Cha; otherwise, the verifier aborts the protocol. (8) It decrypts the challenge Cha to extract $c$ and $h_{RP}$. (9) If $h = h_A$ and $h_A = {\rm hash}(P_A)$, then the protocol run continues as follows. 

(10) Based on the metrics $M_A$, the verifier creates the attestation result $R_{A}$; (11) generates a session key $K_S$; and (12) encrypts $R_A$, $c$, $h_{\rm RP}$, $K_S$ with the key $K_V$ it shares with the RP, resulting in $Res_{\rm RP}$. Next, (13) the verifier encrypts $Res_{\rm RP}$ and $K_S$ using the public key $PK_A$ of the attester, and (14) signs the resulting $Res_A$ using his secret key $SK_V$. The verifier sends $Res_A$ and the signature $\delta_V$ in message (d).

(15) Upon receiving message (d), the attester checks the signature $\delta_V$. (16) If this check succeeds, it decrypts $Res_{A}$ using its secret key $SK_A$, and obtains $Res_A$ and $K_S$ from $Res_A$; otherwise, the attester aborts the protocol. The attester sends $Res_{\rm RP}$ to the RP in message (e).

(17) Upon receiving message (e), the RP decrypts $Res_{\rm RP}$ and obtains $R_A$, $c_{\rm Res}$, $h_{\rm Res}$ and $K_S$. (18) If $c=c_{\rm Res}$ and $h=h_{\rm Res}$, then (19 ) RP will process the attestation result to determine the attester's state; otherwise, it will abort the protocol run. 

Subsequent, application-specific communication between RP and attester (f) depends on the result of step (19); this communication is secured using $K_S$. Please note that before that communication takes place, the RP and attester have not confirmed the possession of $K_S$ to each other.

The requirements for this protocol are as in Section \ref{sec:Requirements}, with the addition of SR5: secrecy of $K_S$. We have verified the security properties SR1-SR5 using ProVerif. Our code is available in GitHub \footnote{\href{https://github.com/Mariam-Dessouki/RAforConstrainedRP/tree/apcr-paper}{\color{blue}{ProVerif Model}}}. 

We remark, first, that an alternative to verifier generating the session key $K_S$ (step (11) in Figure~\ref{fig:verifier-generates-key}), is to generate $K_S$ in the RP, and then send it encrypted with $K_V$ to the verifier in message (a).  
%
Second, the protocol description could be simpler if rather than separately protecting messages (c) and (d), the attester and the verifier would communicate through a secure channel using, e.g., the TLS 1.3 protocol \cite{rfc8446}.

\end{document}